\documentclass[preprint,showpacs,preprintnumbers,amsmath,amssymb]{revtex4}

\usepackage{graphicx}
\usepackage{dcolumn}
\usepackage{bm}

\begin{document}

\preprint{Climate}

\title{Climate Prediction through Statistical Methods}

\author{Bora Akgun}
 \altaffiliation[Presently at ]{Department of Physics, Carnegie Mellon University, Pittsburgh, PA, USA.}
\author{Zeynep Isvan}
 \altaffiliation[Presently at ]{Department of Physics and Astronomy, University Pittsburgh, Pittsburgh, PA, USA.}
\author{Levent Tuter}
 \altaffiliation[Presently at ]{Department of MICROTECHNOLOGY AND NANOSCIENCE, Chalmers University of Technology, Gothenborg, Sweden.}
\author{Mehmet Levent Kurnaz}
 \homepage{http://web.boun.edu.tr/kurnaz}
\affiliation{Physics Department, Bogazici University, Bebek Istanbul Turkey.}%

\date{\today}

\begin{abstract}
Climate change is a reality of today. Paleoclimatic proxies and climate predictions based on coupled atmosphere-ocean general circulation models provide us with temperature data. Using Detrended Fluctuation Analysis, we are investigating the statistical connection between the climate types of the present and these local temperatures. We are relating this issue to some well-known historic climate shifts. Our main result is that the temperature fluctuations with or without a temperature scale attached to them, can be used to classify climates in the absence of other indicators such as pan evaporation and precipitation.

\end{abstract}

\maketitle

\section{\label{sec:level1}Introduction}

Our climate is changing rapidly. We see the effects of this climate shift in every aspect of nature from retreating glaciers \cite{NSIDC1,pelto07} and breaking ice shelves \cite{BBC1,USAT1,NSIDC2} to changing migration patterns of animals \cite{root03,root05} and even the extinction of certain species \cite{thomas04}. While the scientific evidence for a serious climate shift is growing, there are still some groups trying to cloud the public opinion with arguments about past climate changes having little or no effect on our society. The first rudimentary thermometer was invented by Galileo Galilei in 1593. Daniel Gabriel Fahrenheit invented the first mercury thermometer and made it possible for scientists to measure temperatures consistently. Our knowledge about the climates before the time of Galileo and Fahrenheit usually comes from written (or sometimes painted) personal accounts or what kind of wine was produced in that region of the Earth, rather than from irrefutable scientific evidence. Although many such accounts are very useful in determining the climate of the "Old World", we have settled many different parts of the Earth, and hence more information is needed about paleoclimates to make more sense of the changes happening today.

Our scientific efforts are normally based on the analysis of temperature records obtained from various sources like tree rings \cite{fritts76}, ice core data \cite{petit99}, ocean or lake sediments \cite{hill06,gingele07}, or coral reefs \cite{dunbar94}. Since these proxies provide us mostly with a  temperature record, this only gives us limited knowledge of the climate. For example, if we compare Ospina Perez, Colombia to Las Vegas, USA, we can immediately say that the climate of these two cities should be vastly different. Ospina Perez, Colombia has a tropical monsoon climate receiving about 15 times more rain than Las Vegas, USA which is located in a dry/hot low latitude desert. However, if we look at the yearly average temperatures of these two cities, we find the averages to be around 19.1 o C. Therefore, in addition to temperature records, the climatologists are also using average precipitation and evaporation rates to determine the climate of a certain region.

Determining the weather is actually a rather simple issue. A cold day is usually followed by a cold day, and a warm day is usually followed by a warm day. On a larger scale, a colder week is usually followed by a warmer week which corresponds to the average duration of the general weather regimes. But as the longer timescales are governed by different processes like circulation patterns and global warming, defining long-term correlations becomes more difficult. Recently, scaling arguments have been used to analyze climatic data \cite{Kantelhardt13100,Eichner13120,Livina13130,Monetti13140,Bunde13180,Govindan13190,Vjushin13200,Kantelhardt13220,Govindan13230,Koscielny-Bunde13240,Koscielny-Bunde13250,Bodri13500,Bodri13530,Weber2830,Talkner3550,Weber6080,Fraedrich13270,Ivanova14200,Kiely14190,Ivanova14170,
Ivanova14160,Ausloos14150,Ivanova14140,Kitova14130,Ivanova14120,Ivanova14110,Ivanova14100,Ausloos14090,Kurnaz2004a,Kurnaz2004b}.

In order to separate the trends and the correlations we need to eliminate the trends in our temperature data. Several methods are used effectively for this purpose: rescaled range analysis (R/S) \cite{Bodri13500, Bodri13530, Mandelbrot12180, Mandelbrot12190}, wavelet techniques (WT) \cite{Koscielny-Bunde13240, Arneodo5250} and detrended fluctuation analysis (DFA) \cite{Koscielny-Bunde13250, Peng13540}.

Climate change is a major concern for everyone, and it is a focus of atmospheric research. The changes in properties of the cloud cover influence the radiative feedback from the clouds and become one of the important players that effect the climate in general \cite{Ausloos14090}. Considerable effort has been made to study the time scaling of fluctuations of the cloud cover using DFA \cite{Ivanova14200, Kiely14190, Ivanova14170, Ivanova14160, Ivanova14140, Kitova14130, Ivanova14120, Ivanova14110, Ivanova14100}.

Analysis of the temperature fluctuations over a period of decades in different places of the globe has already shown the effectiveness of the application of DFA to characterize the persistence of weather and climate regimes. DFA and WT have been applied to study temperature and precipitation correlations in different climatic zones on the continents and also in the sea-surface temperature of the oceans. The recent results show that the temperatures are long range power law correlated. The long-term persistence of the temperatures can be characterized by an auto-correlation function, $C(n)$, of temperature variations where $n$ is the time between the observations. The auto-correlation function decays as $C(n)\sim n^{-\gamma}$. Though there is some disagreement on the value of $\gamma$, the fact that the persistence of the temperatures can be characterized by this auto-correlation function is firmly established. Different groups have used R/S, DFA and WT analysis \textbf{to show} that $\gamma$ has roughly the same value, $\gamma \simeq 0.7$  for the continental stations \cite{Eichner13120, Bunde13180, Kantelhardt13220, Koscielny-Bunde13250, Bodri13500, Bodri13530, Weber2830, Talkner3550} and roughly $\gamma \simeq 0.4$ for island stations \cite{Eichner13120, Bunde13180} and on the oceans \cite{Monetti13140, Bunde13180}. These methods have also been applied to the temperature predictions of coupled atmosphere-ocean general circulation models \cite{Govindan13190, Fraedrich13270, Fraedrich12720, Blender13260} but there is disagreement on the actual value of the exponent $\gamma$. On one side it is argued that the exponent does not change with the distance from the oceans \cite{Eichner13120} and is roughly $\gamma \simeq 0.7$. On the other side it is said that the scaling exponent is roughly 1 over the oceans, roughly 0.5 over the inner continents and about 0.65 in transition regions \cite{Fraedrich12720}.

Previous work in this area has also shown that there is a slight variation in the scaling exponent between the low-elevation, mountain, continental and maritime stations \cite{Weber2830, Talkner3550, Kurnaz2004b}. Most of these variations are within the standard deviation of $\gamma$ and even though some of the previous work shows a correlation with location and elevation \cite{Weber2830, Talkner3550} it is difficult to establish a general relationship between the statistical nature of the temperature fluctuations and elevation.

In our previous work we have learned that Detrended Fluctuation Analysis of temperature data can be used to suggest a relationship between the temperature fluctuations and climate without any help from precipitation and evaporation data \cite{Kurnaz2004a, Kurnaz2004b}. The present work tries to extend the same method to show that DFA can be used to discover historic climate changes for which no scientific climate records exit. The natural extension of this work will be to quantify present climate changes.

\section{Method}
To remove the seasonal trends that are known to exist in the daily temperature data we need to determine the mean temperature for each day over all the years in the time series. We then calculate the fluctuation of the daily temperature from the mean daily temperature,
\begin{equation}
\Delta T_{i} = T_{i} - <T_{i}>
\end{equation}
where $<T_{i}>$ is the mean daily maximum temperature. Similarly we can also use the mean daily average temperature or the mean daily minimum
temperature in place of the maximum temperature without changing the outcome of the analysis \cite{Talkner3550}. The average temperature for some years can be higher or lower than the average temperature of the time series for the given location as a result of long-term atmospheric processes. To remove such remaining linear trends in the data we applied the Detrended Fluctuation Analysis (DFA) method \cite{Peng13540}, which removes trends in the data to allow for investigation of long-term correlations in the data.

The noise and the nonstationarity in the temperature data usually hinders a reliable and direct calculation of the autocorrelation function $C(n)$.
Instead we calculate a running sum of the temperature fluctuations,
\begin{equation}
y (m) = \sum_{i=1}^n \Delta T_{i}
\end{equation}
where $m = 1,. . . , n$. Next, the time series of the $y(m)$ is divided into nonoverlapping intervals of equal length $n$. In each interval, we fit $y(m)$ to a straight line, $x(m) = km + d$ for each segment and calculate the detrended square variability $F^{2}(n)$ as
\begin{equation}
F^{2}(n) = < \frac{1}{n} \sum_{m = kn + 1}^{(k + 1)n} (y(m) - x(m))^{2}>
\end{equation}
with
\begin{equation}
k = 0, 1, 2, . . . , ( \frac{N}{n} - 1).
\end{equation}

\begin{figure}
\includegraphics[width=12cm]{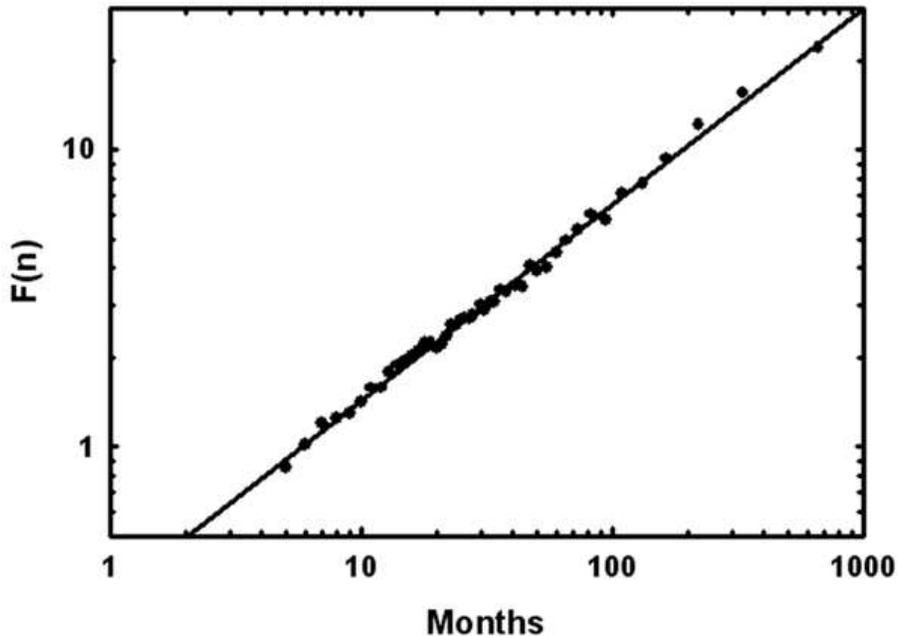}
\caption{\label{fig1} The power law relationship between the time period
and the detrended square variability for New Bedford, MA.}
\end{figure}

If the temperature fluctuations were uncorrelated (white noise) we would expect
\begin{equation}
F(n) \sim n^\alpha
\end{equation}
where  $\alpha = \frac{1}{2}$. If $\alpha > \frac{1}{2}$, we would expect long-range power law correlations in the data for the range of values considered. Fig.~\ref{fig1} shows one example of such an analysis method for New Bedford, MA, USA. The data set for this station spans 100 years of uninterrupted monthly temperature data between the years 1895 and 1994. The data follows a straight line with slope $\alpha = 0.66 \pm 0.01$. Even though there is some scatter in the data after a period of 20 years, the scatter in the data is still within the error estimation of the analysis. This scatter is caused by the fact that the average in the detrended square variability $F^{2}(n)$ has been taken over larger and larger values of $n$, resulting in poor statistics.

\begin{figure}
\includegraphics[width=14cm]{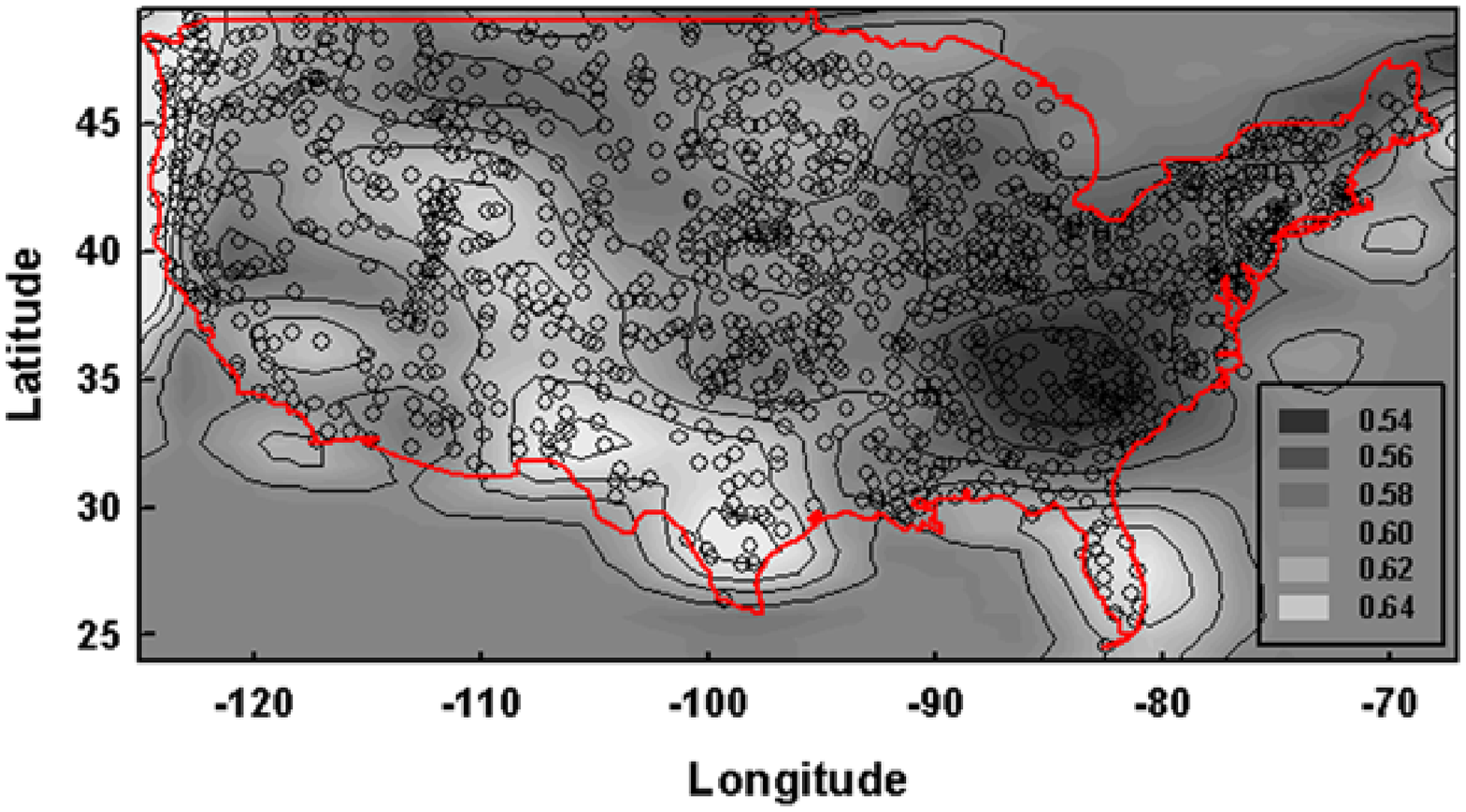}
\caption{\label{fig2} The map of US with power law exponents and the stations marked.}
\end{figure}

Fig.~\ref{fig2} shows a map of the power law exponents for the available 1184 stations in the continental US together with the locations of the stations. This map has been generated from the data using a running average method to generate a grid of 30 × 30 regions over the geographical region. A simple observation about the power law exponents is that even though the power law exponents crowd around the mean value of $\alpha = 0.60 \pm 0.04$, their spatial distribution is not uniform. From the map we can easily distinguish a few regions, namely the northeastern part is darker (lower power law exponent) than the southern and northwestern part (higher power law exponent). As in our previous work \cite{Kurnaz2004b}, we cannot explain this behavior with either the distance from the oceans or the elevation or any simple dependence on geography.

\section{Results}

\begin{figure}
\includegraphics[width=12cm]{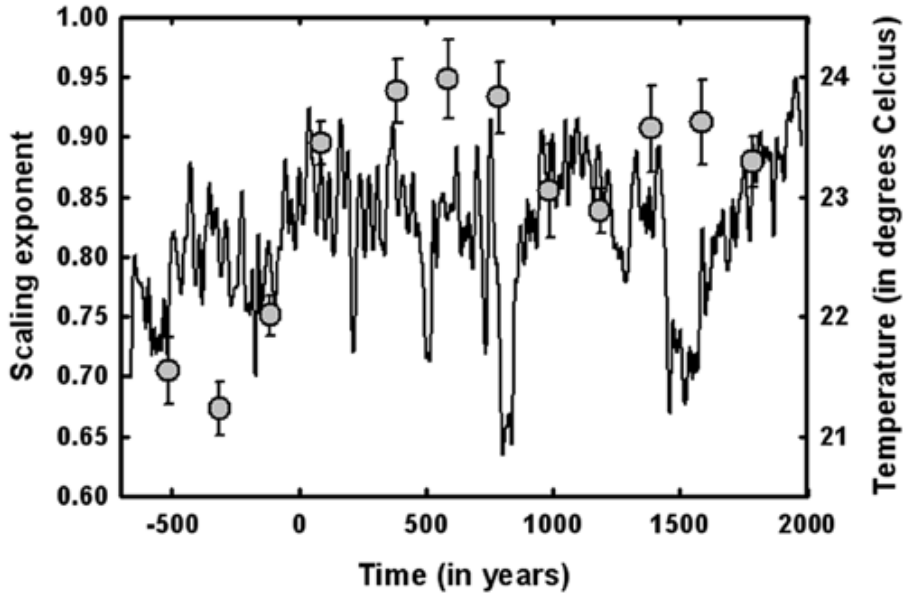}
\caption{\label{fig3} Temperature recontruction from the stalagmite thicknesses in Shihua Cave, Beijing, China (lines). Also plotted on this graph are the scaling exponents obtained from the same data for periods of 400 years (circles).}
\end{figure}

To make some sense of this behavior we took a rather different approach. In the recorded human history there have been quite a few locally important climate changes. Many of these changes have been recorded in the paleoclimatic data. For example, tree ring history from Scandinavia indicates that there was a period of high temperatures between the 9th and 13th centuries, called the "Medieval Warm Period" \cite{Cook13050}. The tree ring temperature history for this period agrees with the glacier data, and \textbf{is} supported by the historical records of Norse seafaring and colonization around the North Atlantic at the end of the 9th century. It is known that during this time the warmer climate helped in producing greater harvests in Iceland and parts of Greenland \cite{Bryson13060} which in turn helped the colonies in Greenland. But when we go to older times, it becomes more and more difficult to relate the paleoclimatic proxies to the climate of a certain region. From paleoclimatic proxies, like tree ring
indices or stacked oxygen-isotope curves derived from deep sea cores, we normally obtain fluctuations of the local temperatures rather than their absolute values. Therefore, it is difficult to construct a method to characterize different climates based solely on the fluctuations of temperature values. As an example of this we have plotted in Fig.~\ref{fig3} the yearly average temperatures reconstructed from stalagmite thicknesses from Shihua Cave near Beijing, China \cite{Tan}. Even though these average temperatures do not show any time dependence, it is widely believed that the climate change in Central Asia around 100 BC was the main reason for the Hsiung-nu expansion towards the west \cite{Chase-Dunn}. Beijing, China is not considered a part of Central Asia, but it is a nearby location for which we have reliable data. When we plot the scaling exponents obtained over 400 year periods for this region together with the temperature data, we see that scaling exponents exhibit a jump around 100 BC whereas the temperatures stay fairly constant (Fig.~\ref{fig3}). This leads us to believe that temperature fluctuations, together with the scaling exponents, can be used as climate indicators in the absence of other indicators like pan evaporation or precipitation.

\begin{figure}
\includegraphics[width=12cm]{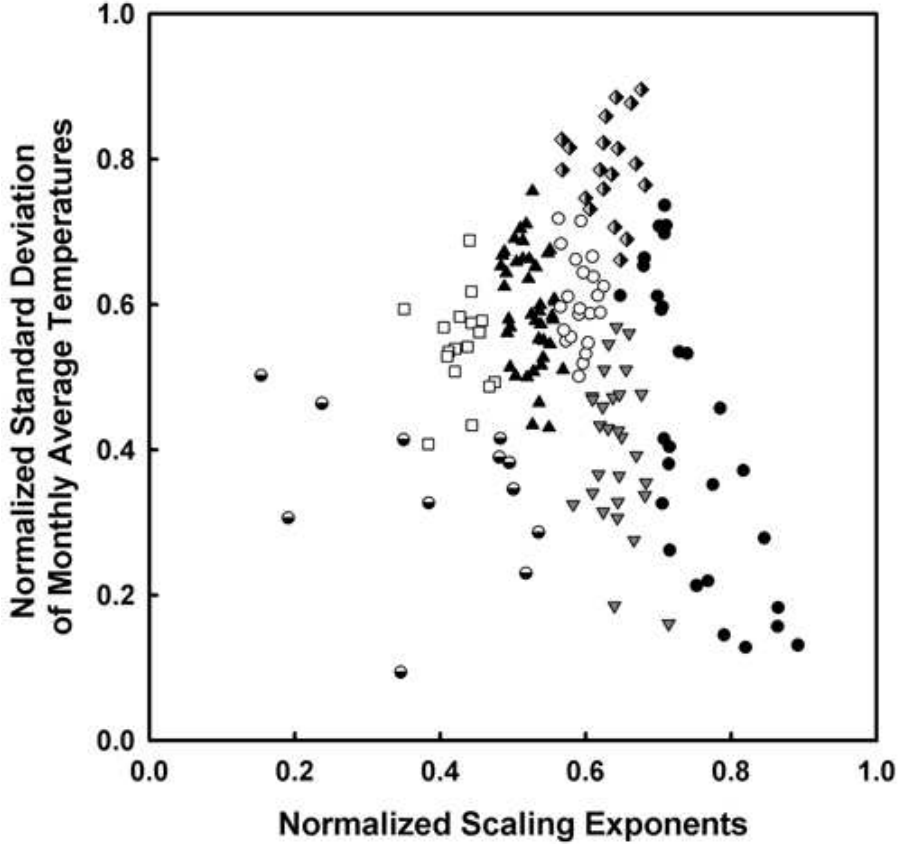}
\caption{\label{fig4} The scaling exponents plotted against the standard deviation of the temperature fluctuations for different climate types. Examples: Black circles - Pensacola FL, black triangles - Ann Arbor MI, white squares - Minneapolis MN, Half-filled circles - Boulder CO, Grey triangles - San Antonio TX, Half-filled diamonds - Blacksburg VA, white circles - Stillwater OK}
\end{figure}

The Koeppen climate classification\cite{Koeppen13800} is one of the most widely used climate classification systems. Although this system is based on the concept that native vegetation expresses the climate, it combines average annual and monthly temperatures and precipitation, and the seasonality of precipitation to classify climates. According to this classification scheme, there are about 14 different climate classes in the continental US. These classes range from tropical monsoon climate (Am) to hemiboreal climates (Dfb, Dsb, Dwb). Some of these classes are similar to each other in their characteristics but there is criticism because some regions are very broad. For example, Florida and New Jersey are both in the class Cfa (humid subtropical). In our work, our basic assumption is that the temperature fluctuations may be used to classify climates in the absence of other indicators. Therefore, we divided the climate of the US into distinct groups. These groups have tenuous connections to Koeppen classes. We then plotted the scaling exponents versus the standard deviation of the temperature fluctuations for locations fitting into these climate classes. Fig.~\ref{fig4} shows these sample locations. As we can see from this graph, locations with similar climates tend to have similar scaling exponent/standard deviation values.

We must be clear about one point: The reason why this method is being utilized is not to find an alternate classification scheme to already existing ones. The basic question to be answered is the following: When we have the temperature data from a paleoclimatic proxy, we know what the temperatures are, but we have little information about the climate. Can we make sense out of the temperature fluctuations from that location? Is it more like Ospina Perez, Colombia or closer to Las Vegas, NV? If we only have the standard deviations of the temperature fluctuations and the scaling exponents resulting from those distributions, obtaining clusters belonging to different climates would be extremely difficult. However, we possess additional information in the location of the weather stations. Therefore our task is not finding appropriate clusters, but classifying new data into already existing clusters. To aid us in classification and to support the claim that the differences in the scaling exponents of different geographical/climatic regions are statistically significant we have used the support vector machine (SVM) algorithm \cite{Vapnik13740} for data classification.

The inputs for a regular data classification task are training and testing data. The training dataset consists of one target value (in our case belonging to a specific climate type) and several features (like the standard deviation of the temperature fluctuations and the scaling exponent). We also need to tell the algorithm what we do not want. In the training set we have used 10 positive target values (belonging to the climate class we are analyzing) and 20 negative target points (belonging to different climate classes). The aim of SVM \cite{Joachims13750} is to produce a model which then predicts the target value of data instances in the test set. In the test set, we supplied the algorithm with all of the stations in our dataset and asked the program to identify the different climate types. The performance of such an algorithm is usually quantified by its accuracy during the test phase which mainly depends on the correct treatment of true positives (TP) and true negatives (TN). It is also important to distinguish between two types of errors: A false positive (FP) and a false negative (FN). Consequently, the performance of the prediction is better judged if we add two more quantifiers, sensitivity and specificity. The accuracy of the data classification is defined as the ratio between the number of correctly identified samples and the total number of samples:

\begin{equation}
accuracy = \frac{TP+TN}{TP+FP+TN+FN}
\end{equation}

The sensitivity is the ratio between the number of true positive predictions and the number of positive instances in the test set:

\begin{equation}
sensitivity = \frac{TP}{TP+FN}
\end{equation}

Finally, the specificity is defined as the ratio between the number of true negative predictions and the number of negative instances in the test set:

\begin{equation}
specificity = \frac{TN}{FP+TN}
\end{equation}

For our analysis, we used the different climates mentioned above in Fig.~\ref{fig4}. The results are summarized in Table~\ref{tablesums}. In this table we see that our ability to predict accurately, sensitively and specifically never goes below 96\%. However at this point we must explain why we came up with seven different climate classes. If we run the same classification algorithm with more than seven classes, the accuracy, or the sensitivity, or the specificity goes below 95 \%. Conversely, we can get even better accuracy/sensitivity/specificity by decreasing the number of classes. Therefore, we restricted ourselves to identifying seven different climatic regions. When dealing with paleoclimatic data, the accuracy/sensitivity/specificity levels can be adjusted to the needs of the situation.

\begin{table}
\caption{SVM analysis of the 1184 weather stations in different
climate zones. Accuracy, sensitivity, and specificity are defined
in the text.}
\label{tablesums}
\begin{ruledtabular}
\begin{tabular}{lccc}
Climate Type & Accuracy & Sensitivity & Specificity \\
\hline Type 1 (Pensacola FL) & $99.7 \%$ & $100.0 \%$ & $99.7 \%$ \\
Type 2 (Ann Arbor MI) &$98.1 \%$ & $96.0 \%$ & $99.3 \%$ \\
Type 3 (Minneapolis MN) & $98.8 \%$ & $99.3 \%$ & $98.7 \%$ \\
Type 4 (Boulder CO) & $99.3 \%$ & $100.0 \%$ & $99.3 \%$ \\
Type 5 (San Antonio TX) & $97.8 \%$ & $99.3 \%$ & $97.6 \%$ \\
Type 6 (Blacksburg VA) & $98.6 \%$ & $100.0 \%$ & $98.4 \%$ \\
Type 7 (Stillwater OK) & $96.0 \%$ & $96.4 \%$ & $96.0 \%$ \\
\end{tabular}
\end{ruledtabular}
\end{table}

\begin{figure}
\includegraphics[width=12cm]{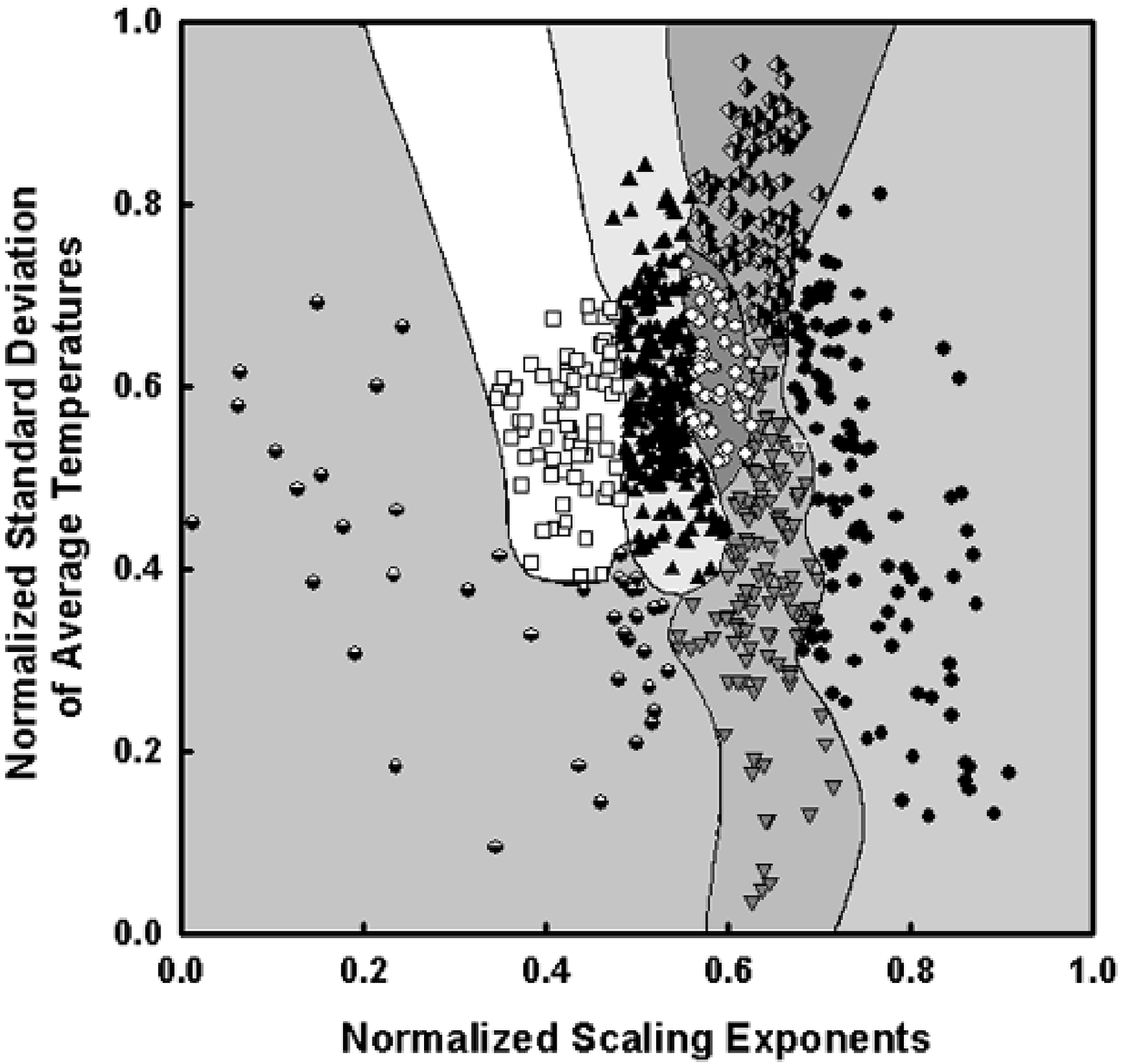}
\caption{\label{fig5} The scaling exponents plotted against the standard deviation of the temperature fluctuations for different climate types after the classification by the SVM algorithm. Examples: Black circles - Type 1, black triangles - Type 2, white squares - Type 3, Half-filled circles - Type 4, Grey triangles - Type 5, Half-filled diamonds - Type 6, white circles - Type 7}
\end{figure}

\begin{figure}
\includegraphics[width=14cm]{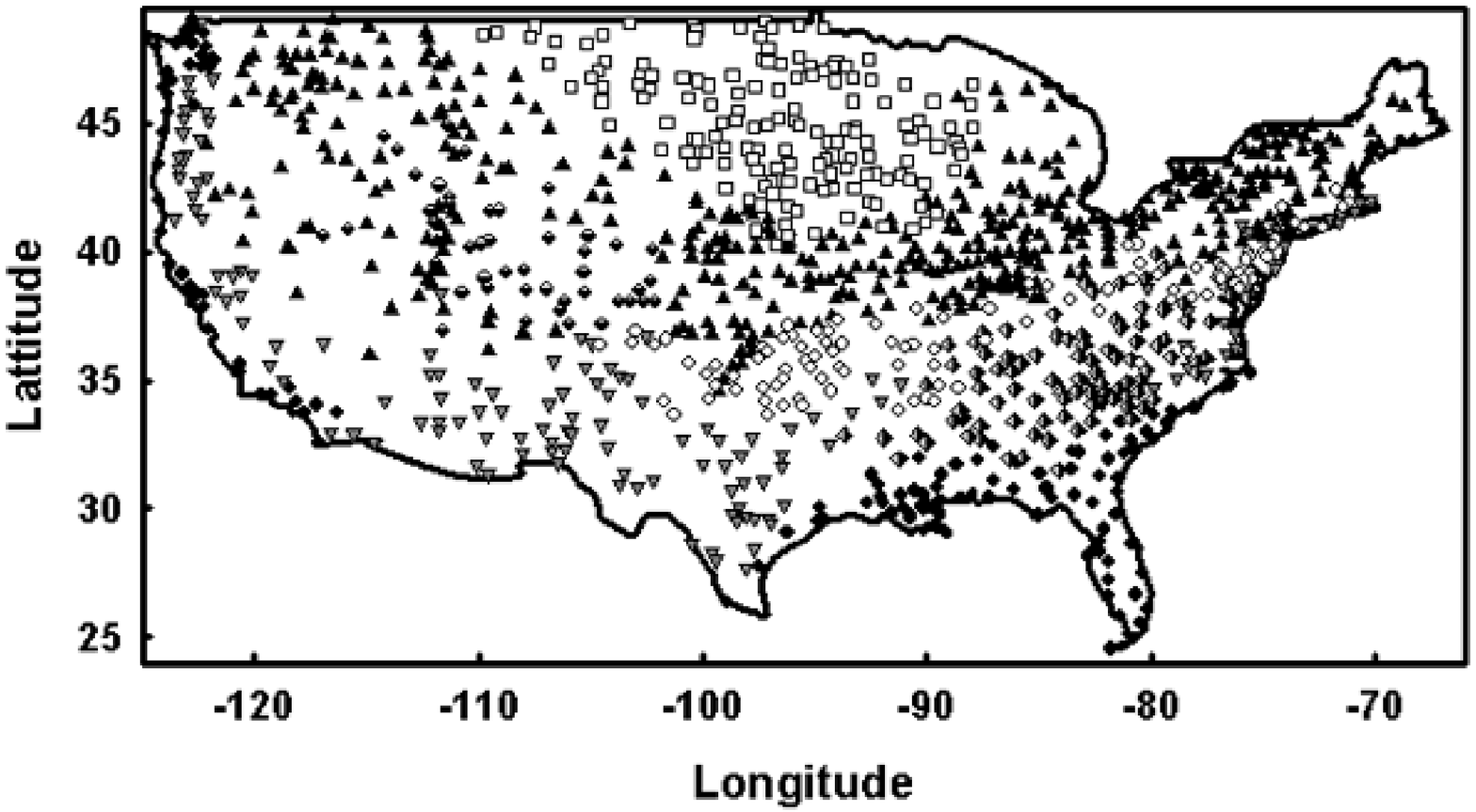}
\caption{\label{fig6} The map of US with the classification results for different climate types and the stations marked. Examples: Black circles - Type 1, black triangles - Type 2, white squares - Type 3, Half-filled circles - Type 4, Grey triangles - Type 5, Half-filled diamonds - Type 6, white circles - Type 7}
\end{figure}

In Fig.~\ref{fig5} we show the result for all the available US data. Finally we show these locations on the US map (Fig.~\ref{fig6}). As can be expected, the result is not always perfect, however, we should keep in mind that these results have been obtained without other indicators such as precipitation or evaporation. The addition of such parameters into the same analysis system may make this tool a better categorization method, however as that area has been contested by many climatologists for more than a hundred years, we have restricted our analysis and results to the classification problem at hand. 

\section{Conclusion}

We have investigated the power law behavior of the temperature fluctuations to gain insight into a statistical description of different climate types. Normally we need other indicators such as the pan evaporation and precipitation to classify the climates, but two different fields, both connected with the global climate change, led us to this approach. First, different projections for the future of the climate change give us valuable data about the future temperatures and possibly precipitation. And second, paleoclimatology gives us different proxies to determine the temperatures of ancient times. However, as both of these areas are focused on determining the temperatures, our aim is to aid this area of research by developing a tool that uses temperature fluctuations to determine the climate.

Our results suggest that on the basis of the power law scaling of the temperature fluctuations we can distinguish between different climates. This is by no means a new way to classify climates. It is only intended to compare the climate of a spatio-temporal location for which we only have temperature fluctuations without the exact units to the present day climate for which we possess other meteorological measurements in addition to temperature. The real challenge comes when we try to expand this
analysis into the past, since, to our knowledge, no reliable monthly data exists beyond 218 years ago \cite{Koscielny-Bunde13240}. However, previous work where authors have established long-range power law behavior using rescaled-range analysis \cite{Fluegeman9510, Wang7920} gives us hope about expanding the use of this method we have developed.

The climate projections for the future are crucial in our understanding of how climate change impacts the environment. When we use these global climate models to extrapolate the climate changes into the future, the method of Detrended Fluctuation Analysis proves to be a powerful tool \cite{Govindan13190, Fraedrich13270}. The present work may help to fill in the voids in existing climate models by providing further information on climate and temperature while improvements are made on the length scales these climate models cover along with their applicability to future timescales.

\begin{acknowledgments}
Data was provided by U.S. Historical Climatology Network and the National Climatic Data Center.
\end{acknowledgments}


\begin{thebibliography}{58}
\expandafter\ifx\csname natexlab\endcsname\relax\def\natexlab#1{#1}\fi
\expandafter\ifx\csname bibnamefont\endcsname\relax
  \def\bibnamefont#1{#1}\fi
\expandafter\ifx\csname bibfnamefont\endcsname\relax
  \def\bibfnamefont#1{#1}\fi
\expandafter\ifx\csname citenamefont\endcsname\relax
  \def\citenamefont#1{#1}\fi
\expandafter\ifx\csname url\endcsname\relax
  \def\url#1{\texttt{#1}}\fi
\expandafter\ifx\csname urlprefix\endcsname\relax\def\urlprefix{URL }\fi
\providecommand{\bibinfo}[2]{#2}
\providecommand{\eprint}[2][]{\url{#2}}

\bibitem[{NSI(2006)}]{NSIDC1}
\emph{\bibinfo{title}{Global Glacier Recession}},
  \bibinfo{organization}{National Snow and Ice Data Center}
  (\bibinfo{year}{2006}),
  \bibinfo{note}{http://nsidc.org/glims/glaciermelt/index.html}.

\bibitem[{\citenamefont{Pelto}(2007)}]{pelto07}
\bibinfo{author}{\bibfnamefont{M.~S.} \bibnamefont{Pelto}},
  \bibinfo{journal}{The Cryosphere Discuss.} \textbf{\bibinfo{volume}{1}},
  \bibinfo{pages}{17} (\bibinfo{year}{2007}).

\bibitem[{BBC(2002)}]{BBC1}
\emph{\bibinfo{title}{Antarctic ice shelf breaks apart}},
  \bibinfo{organization}{BBC News} (\bibinfo{year}{2002}),
  \bibinfo{note}{http://news.bbc.co.uk/2/ hi/science/nature/1880566.stm}.

\bibitem[{USA(2003)}]{USAT1}
\emph{\bibinfo{title}{Ice shelf in warmest part of Antarctica collapses}},
  \bibinfo{organization}{USA Today} (\bibinfo{year}{2003}),
  \bibinfo{note}{http://www.usatoday.com/news/science/cold-science/2003-03-19-%
ant-ice.htm}.

\bibitem[{NSI(2002)}]{NSIDC2}
\emph{\bibinfo{title}{Larsen B Ice Shelf Collapses in Antarctica}},
  \bibinfo{organization}{National Snow and Ice Data Center}
  (\bibinfo{year}{2002}),
  \bibinfo{note}{http://nsidc.org/iceshelves/larsenb2002/}.

\bibitem[{\citenamefont{Root et~al.}(2003)\citenamefont{Root, Price, Hall,
  Schneider, Rosenzweig, and Pounds}}]{root03}
\bibinfo{author}{\bibfnamefont{T.~L.} \bibnamefont{Root}},
  \bibinfo{author}{\bibfnamefont{J.~T.} \bibnamefont{Price}},
  \bibinfo{author}{\bibfnamefont{K.~R.} \bibnamefont{Hall}},
  \bibinfo{author}{\bibfnamefont{S.~H.} \bibnamefont{Schneider}},
  \bibinfo{author}{\bibfnamefont{C.}~\bibnamefont{Rosenzweig}},
  \bibnamefont{and} \bibinfo{author}{\bibfnamefont{A.}~\bibnamefont{Pounds}},
  \bibinfo{journal}{Nature} \textbf{\bibinfo{volume}{421}},
  \bibinfo{pages}{6918} (\bibinfo{year}{2003}).

\bibitem[{\citenamefont{Root et~al.}(2005)\citenamefont{Root, MacMynowski,
  Mastrandrea, and Schneider}}]{root05}
\bibinfo{author}{\bibfnamefont{T.~L.} \bibnamefont{Root}},
  \bibinfo{author}{\bibfnamefont{D.}~\bibnamefont{MacMynowski}},
  \bibinfo{author}{\bibfnamefont{M.~D.} \bibnamefont{Mastrandrea}},
  \bibnamefont{and} \bibinfo{author}{\bibfnamefont{S.~H.}
  \bibnamefont{Schneider}}, \bibinfo{journal}{PNAS}
  \textbf{\bibinfo{volume}{102}}, \bibinfo{pages}{7465} (\bibinfo{year}{2005}).

\bibitem[{\citenamefont{Thomas et~al.}(2004)\citenamefont{Thomas, Cameron,
  Green, Bakkenes, Beaumont, Collingham, Erasmus, de~Siqueira, Grainger, Hannah
  et~al.}}]{thomas04}
\bibinfo{author}{\bibfnamefont{C.~D.} \bibnamefont{Thomas}},
  \bibinfo{author}{\bibfnamefont{A.}~\bibnamefont{Cameron}},
  \bibinfo{author}{\bibfnamefont{R.~E.} \bibnamefont{Green}},
  \bibinfo{author}{\bibfnamefont{M.}~\bibnamefont{Bakkenes}},
  \bibinfo{author}{\bibfnamefont{L.~J.} \bibnamefont{Beaumont}},
  \bibinfo{author}{\bibfnamefont{Y.~C.} \bibnamefont{Collingham}},
  \bibinfo{author}{\bibfnamefont{B.~F.~N.} \bibnamefont{Erasmus}},
  \bibinfo{author}{\bibfnamefont{M.~F.} \bibnamefont{de~Siqueira}},
  \bibinfo{author}{\bibfnamefont{A.}~\bibnamefont{Grainger}},
  \bibinfo{author}{\bibfnamefont{L.}~\bibnamefont{Hannah}},
  \bibnamefont{et~al.}, \bibinfo{journal}{Nature}
  \textbf{\bibinfo{volume}{427}}, \bibinfo{pages}{145} (\bibinfo{year}{2004}).

\bibitem[{\citenamefont{Fritts}(1976)}]{fritts76}
\bibinfo{author}{\bibfnamefont{H.~C.} \bibnamefont{Fritts}},
  \emph{\bibinfo{title}{Tree Rings and Climate}} (\bibinfo{publisher}{Academic
  Press}, \bibinfo{year}{1976}).

\bibitem[{\citenamefont{Petit et~al.}(1999)\citenamefont{Petit, Jouzel,
  Raynaud, Barkov, Barnola, Basile, Bender, Chappellaz, Davis, Delaygue
  et~al.}}]{petit99}
\bibinfo{author}{\bibfnamefont{J.~R.} \bibnamefont{Petit}},
  \bibinfo{author}{\bibfnamefont{J.}~\bibnamefont{Jouzel}},
  \bibinfo{author}{\bibfnamefont{D.}~\bibnamefont{Raynaud}},
  \bibinfo{author}{\bibfnamefont{N.~I.} \bibnamefont{Barkov}},
  \bibinfo{author}{\bibfnamefont{J.~M.} \bibnamefont{Barnola}},
  \bibinfo{author}{\bibfnamefont{I.}~\bibnamefont{Basile}},
  \bibinfo{author}{\bibfnamefont{M.}~\bibnamefont{Bender}},
  \bibinfo{author}{\bibfnamefont{J.}~\bibnamefont{Chappellaz}},
  \bibinfo{author}{\bibfnamefont{J.}~\bibnamefont{Davis}},
  \bibinfo{author}{\bibfnamefont{G.}~\bibnamefont{Delaygue}},
  \bibnamefont{et~al.}, \bibinfo{journal}{Nature}
  \textbf{\bibinfo{volume}{399}}, \bibinfo{pages}{429} (\bibinfo{year}{1999}).

\bibitem[{\citenamefont{Hill et~al.}(2006)\citenamefont{Hill, Kennett, Pak,
  Behl, Robert, and Beaufort}}]{hill06}
\bibinfo{author}{\bibfnamefont{T.~M.} \bibnamefont{Hill}},
  \bibinfo{author}{\bibfnamefont{J.~P.} \bibnamefont{Kennett}},
  \bibinfo{author}{\bibfnamefont{D.~K.} \bibnamefont{Pak}},
  \bibinfo{author}{\bibfnamefont{R.}~\bibnamefont{Behl}},
  \bibinfo{author}{\bibfnamefont{C.}~\bibnamefont{Robert}}, \bibnamefont{and}
  \bibinfo{author}{\bibfnamefont{L.}~\bibnamefont{Beaufort}},
  \bibinfo{journal}{Quaternary Science Reviews} \textbf{\bibinfo{volume}{25}},
  \bibinfo{pages}{2835} (\bibinfo{year}{2006}).

\bibitem[{\citenamefont{Gingele et~al.}(2007)\citenamefont{Gingele, de~Deckker,
  and Norman}}]{gingele07}
\bibinfo{author}{\bibfnamefont{F.}~\bibnamefont{Gingele}},
  \bibinfo{author}{\bibfnamefont{P.}~\bibnamefont{de~Deckker}},
  \bibnamefont{and} \bibinfo{author}{\bibfnamefont{M.}~\bibnamefont{Norman}},
  \bibinfo{journal}{Earth and Planetary Sci. Lett.}
  \textbf{\bibinfo{volume}{255}}, \bibinfo{pages}{257} (\bibinfo{year}{2007}).

\bibitem[{\citenamefont{Dunbar et~al.}(1994)\citenamefont{Dunbar, Wellington,
  Colgan, and Glynn}}]{dunbar94}
\bibinfo{author}{\bibfnamefont{R.~B.} \bibnamefont{Dunbar}},
  \bibinfo{author}{\bibfnamefont{G.}~\bibnamefont{Wellington}},
  \bibinfo{author}{\bibfnamefont{M.}~\bibnamefont{Colgan}}, \bibnamefont{and}
  \bibinfo{author}{\bibfnamefont{P.}~\bibnamefont{Glynn}},
  \bibinfo{journal}{Paleooceonography} \textbf{\bibinfo{volume}{9}},
  \bibinfo{pages}{291} (\bibinfo{year}{1994}).

\bibitem[{\citenamefont{Kantelhardt et~al.}(2003)\citenamefont{Kantelhardt,
  Rybski, Zschiegner, Braun, Koscielny-Bunde, Livina, Havlin, and
  Bunde}}]{Kantelhardt13100}
\bibinfo{author}{\bibfnamefont{J.~W.} \bibnamefont{Kantelhardt}},
  \bibinfo{author}{\bibfnamefont{D.}~\bibnamefont{Rybski}},
  \bibinfo{author}{\bibfnamefont{S.}~\bibnamefont{Zschiegner}},
  \bibinfo{author}{\bibfnamefont{P.}~\bibnamefont{Braun}},
  \bibinfo{author}{\bibfnamefont{E.}~\bibnamefont{Koscielny-Bunde}},
  \bibinfo{author}{\bibfnamefont{V.}~\bibnamefont{Livina}},
  \bibinfo{author}{\bibfnamefont{S.}~\bibnamefont{Havlin}}, \bibnamefont{and}
  \bibinfo{author}{\bibfnamefont{A.}~\bibnamefont{Bunde}},
  \bibinfo{journal}{Physica A} \textbf{\bibinfo{volume}{330}},
  \bibinfo{pages}{240} (\bibinfo{year}{2003}).

\bibitem[{\citenamefont{Eichner et~al.}(2003)\citenamefont{Eichner,
  Koscielny-Bunde, Bunde, Havlin, and Schnellnhuber}}]{Eichner13120}
\bibinfo{author}{\bibfnamefont{J.}~\bibnamefont{Eichner}},
  \bibinfo{author}{\bibfnamefont{E.}~\bibnamefont{Koscielny-Bunde}},
  \bibinfo{author}{\bibfnamefont{A.}~\bibnamefont{Bunde}},
  \bibinfo{author}{\bibfnamefont{S.}~\bibnamefont{Havlin}}, \bibnamefont{and}
  \bibinfo{author}{\bibfnamefont{H.~J.} \bibnamefont{Schnellnhuber}},
  \bibinfo{journal}{Phys. Rev. E} \textbf{\bibinfo{volume}{68}},
  \bibinfo{pages}{046133} (\bibinfo{year}{2003}).

\bibitem[{\citenamefont{Livina et~al.}(2003)\citenamefont{Livina, Ashkenazy,
  Braun, Monetti, Bunde, and Havlin}}]{Livina13130}
\bibinfo{author}{\bibfnamefont{V.}~\bibnamefont{Livina}},
  \bibinfo{author}{\bibfnamefont{Y.}~\bibnamefont{Ashkenazy}},
  \bibinfo{author}{\bibfnamefont{P.}~\bibnamefont{Braun}},
  \bibinfo{author}{\bibfnamefont{R.~A.} \bibnamefont{Monetti}},
  \bibinfo{author}{\bibfnamefont{A.}~\bibnamefont{Bunde}}, \bibnamefont{and}
  \bibinfo{author}{\bibfnamefont{S.}~\bibnamefont{Havlin}},
  \bibinfo{journal}{Phys. Rev. E} \textbf{\bibinfo{volume}{67}},
  \bibinfo{pages}{042101} (\bibinfo{year}{2003}).

\bibitem[{\citenamefont{Monetti et~al.}(2003)\citenamefont{Monetti, Havlin, and
  Bunde}}]{Monetti13140}
\bibinfo{author}{\bibfnamefont{R.~A.} \bibnamefont{Monetti}},
  \bibinfo{author}{\bibfnamefont{S.}~\bibnamefont{Havlin}}, \bibnamefont{and}
  \bibinfo{author}{\bibfnamefont{A.}~\bibnamefont{Bunde}},
  \bibinfo{journal}{Physica A} \textbf{\bibinfo{volume}{320}},
  \bibinfo{pages}{581} (\bibinfo{year}{2003}).

\bibitem[{\citenamefont{Bunde and Havlin}(2002)}]{Bunde13180}
\bibinfo{author}{\bibfnamefont{A.}~\bibnamefont{Bunde}} \bibnamefont{and}
  \bibinfo{author}{\bibfnamefont{S.}~\bibnamefont{Havlin}},
  \bibinfo{journal}{Physica A} \textbf{\bibinfo{volume}{314}},
  \bibinfo{pages}{15} (\bibinfo{year}{2002}).

\bibitem[{\citenamefont{Govindan et~al.}(2002)\citenamefont{Govindan, Vjushin,
  Bunde, Brenner, Ashkenazy, Havlin, and Schnellnhuber}}]{Govindan13190}
\bibinfo{author}{\bibfnamefont{R.}~\bibnamefont{Govindan}},
  \bibinfo{author}{\bibfnamefont{D.}~\bibnamefont{Vjushin}},
  \bibinfo{author}{\bibfnamefont{A.}~\bibnamefont{Bunde}},
  \bibinfo{author}{\bibfnamefont{S.}~\bibnamefont{Brenner}},
  \bibinfo{author}{\bibfnamefont{Y.}~\bibnamefont{Ashkenazy}},
  \bibinfo{author}{\bibfnamefont{S.}~\bibnamefont{Havlin}}, \bibnamefont{and}
  \bibinfo{author}{\bibfnamefont{H.~J.} \bibnamefont{Schnellnhuber}},
  \bibinfo{journal}{Phys. Rev. Lett.} \textbf{\bibinfo{volume}{89}},
  \bibinfo{pages}{028501} (\bibinfo{year}{2002}).

\bibitem[{\citenamefont{Vjushin et~al.}(2002)\citenamefont{Vjushin, Govindan,
  Brenner, Bunde, Brenner, Havlin, and Schnellnhuber}}]{Vjushin13200}
\bibinfo{author}{\bibfnamefont{D.}~\bibnamefont{Vjushin}},
  \bibinfo{author}{\bibfnamefont{R.}~\bibnamefont{Govindan}},
  \bibinfo{author}{\bibfnamefont{S.}~\bibnamefont{Brenner}},
  \bibinfo{author}{\bibfnamefont{A.}~\bibnamefont{Bunde}},
  \bibinfo{author}{\bibfnamefont{S.}~\bibnamefont{Brenner}},
  \bibinfo{author}{\bibfnamefont{S.}~\bibnamefont{Havlin}}, \bibnamefont{and}
  \bibinfo{author}{\bibfnamefont{H.~J.} \bibnamefont{Schnellnhuber}},
  \bibinfo{journal}{J. Phys.: Cond. Matt.} \textbf{\bibinfo{volume}{14}},
  \bibinfo{pages}{2275} (\bibinfo{year}{2002}).

\bibitem[{\citenamefont{Kantelhardt et~al.}(2001)\citenamefont{Kantelhardt,
  Koscielny-Bunde, Rego, Havlin, and Bunde}}]{Kantelhardt13220}
\bibinfo{author}{\bibfnamefont{J.~W.} \bibnamefont{Kantelhardt}},
  \bibinfo{author}{\bibfnamefont{E.}~\bibnamefont{Koscielny-Bunde}},
  \bibinfo{author}{\bibfnamefont{H.~H.~A.} \bibnamefont{Rego}},
  \bibinfo{author}{\bibfnamefont{S.}~\bibnamefont{Havlin}}, \bibnamefont{and}
  \bibinfo{author}{\bibfnamefont{A.}~\bibnamefont{Bunde}},
  \bibinfo{journal}{Physica A} \textbf{\bibinfo{volume}{295}},
  \bibinfo{pages}{441} (\bibinfo{year}{2001}).

\bibitem[{\citenamefont{Govindan et~al.}(2001)\citenamefont{Govindan, Vjushin,
  Brenner, Bunde, Havlin, and Schnellnhuber}}]{Govindan13230}
\bibinfo{author}{\bibfnamefont{R.}~\bibnamefont{Govindan}},
  \bibinfo{author}{\bibfnamefont{D.}~\bibnamefont{Vjushin}},
  \bibinfo{author}{\bibfnamefont{S.}~\bibnamefont{Brenner}},
  \bibinfo{author}{\bibfnamefont{A.}~\bibnamefont{Bunde}},
  \bibinfo{author}{\bibfnamefont{S.}~\bibnamefont{Havlin}}, \bibnamefont{and}
  \bibinfo{author}{\bibfnamefont{H.~J.} \bibnamefont{Schnellnhuber}},
  \bibinfo{journal}{Physica A} \textbf{\bibinfo{volume}{294}},
  \bibinfo{pages}{239} (\bibinfo{year}{2001}).

\bibitem[{\citenamefont{Koscielny-Bunde
  et~al.}(1998)\citenamefont{Koscielny-Bunde, Bunde, Havlin, Roman, Goldreich,
  and Schnellnhuber}}]{Koscielny-Bunde13240}
\bibinfo{author}{\bibfnamefont{E.}~\bibnamefont{Koscielny-Bunde}},
  \bibinfo{author}{\bibfnamefont{A.}~\bibnamefont{Bunde}},
  \bibinfo{author}{\bibfnamefont{S.}~\bibnamefont{Havlin}},
  \bibinfo{author}{\bibfnamefont{H.~E.} \bibnamefont{Roman}},
  \bibinfo{author}{\bibfnamefont{Y.}~\bibnamefont{Goldreich}},
  \bibnamefont{and} \bibinfo{author}{\bibfnamefont{H.~J.}
  \bibnamefont{Schnellnhuber}}, \bibinfo{journal}{Phys. Rev. Lett.}
  \textbf{\bibinfo{volume}{81}}, \bibinfo{pages}{729} (\bibinfo{year}{1998}).

\bibitem[{\citenamefont{Koscielny-Bunde
  et~al.}(1996)\citenamefont{Koscielny-Bunde, Bunde, Havlin, and
  Goldreich}}]{Koscielny-Bunde13250}
\bibinfo{author}{\bibfnamefont{E.}~\bibnamefont{Koscielny-Bunde}},
  \bibinfo{author}{\bibfnamefont{A.}~\bibnamefont{Bunde}},
  \bibinfo{author}{\bibfnamefont{S.}~\bibnamefont{Havlin}}, \bibnamefont{and}
  \bibinfo{author}{\bibfnamefont{Y.}~\bibnamefont{Goldreich}},
  \bibinfo{journal}{Physica A} \textbf{\bibinfo{volume}{231}},
  \bibinfo{pages}{231} (\bibinfo{year}{1996}).

\bibitem[{\citenamefont{Bodri}(1995)}]{Bodri13500}
\bibinfo{author}{\bibfnamefont{L.}~\bibnamefont{Bodri}},
  \bibinfo{journal}{Theoretical and Applied Climatology}
  \textbf{\bibinfo{volume}{51}}, \bibinfo{pages}{51} (\bibinfo{year}{1995}).

\bibitem[{\citenamefont{Bodri}(1994)}]{Bodri13530}
\bibinfo{author}{\bibfnamefont{L.}~\bibnamefont{Bodri}},
  \bibinfo{journal}{Theoretical and Applied Climatology}
  \textbf{\bibinfo{volume}{49}}, \bibinfo{pages}{53} (\bibinfo{year}{1994}).

\bibitem[{\citenamefont{Weber and Talkner}(2001)}]{Weber2830}
\bibinfo{author}{\bibfnamefont{R.~O.} \bibnamefont{Weber}} \bibnamefont{and}
  \bibinfo{author}{\bibfnamefont{P.}~\bibnamefont{Talkner}},
  \bibinfo{journal}{J. Geophys. Res. Atmos.} \textbf{\bibinfo{volume}{106}},
  \bibinfo{pages}{20131} (\bibinfo{year}{2001}).

\bibitem[{\citenamefont{Talkner and Weber}(2000)}]{Talkner3550}
\bibinfo{author}{\bibfnamefont{P.}~\bibnamefont{Talkner}} \bibnamefont{and}
  \bibinfo{author}{\bibfnamefont{R.~O.} \bibnamefont{Weber}},
  \bibinfo{journal}{Phys. Rev. E} \textbf{\bibinfo{volume}{62}},
  \bibinfo{pages}{150} (\bibinfo{year}{2000}).

\bibitem[{\citenamefont{Weber et~al.}(1994)\citenamefont{Weber, Talkner, and
  Stefanicki}}]{Weber6080}
\bibinfo{author}{\bibfnamefont{R.~O.} \bibnamefont{Weber}},
  \bibinfo{author}{\bibfnamefont{P.}~\bibnamefont{Talkner}}, \bibnamefont{and}
  \bibinfo{author}{\bibfnamefont{G.}~\bibnamefont{Stefanicki}},
  \bibinfo{journal}{Geophys. Res. Lett.} \textbf{\bibinfo{volume}{21}},
  \bibinfo{pages}{673} (\bibinfo{year}{1994}).

\bibitem[{\citenamefont{Fraedrich and Blender}(2003)}]{Fraedrich13270}
\bibinfo{author}{\bibfnamefont{K.}~\bibnamefont{Fraedrich}} \bibnamefont{and}
  \bibinfo{author}{\bibfnamefont{R.}~\bibnamefont{Blender}},
  \bibinfo{journal}{Phys. Rev. Lett.} \textbf{\bibinfo{volume}{90}},
  \bibinfo{pages}{108501} (\bibinfo{year}{2003}).

\bibitem[{\citenamefont{Ivanova and Ackermann}(1999)}]{Ivanova14200}
\bibinfo{author}{\bibfnamefont{K.}~\bibnamefont{Ivanova}} \bibnamefont{and}
  \bibinfo{author}{\bibfnamefont{T.}~\bibnamefont{Ackermann}},
  \bibinfo{journal}{Phys. Rev. E} \textbf{\bibinfo{volume}{59}},
  \bibinfo{pages}{2778} (\bibinfo{year}{1999}).

\bibitem[{\citenamefont{Kiely and Ivanova}(1999)}]{Kiely14190}
\bibinfo{author}{\bibfnamefont{G.}~\bibnamefont{Kiely}} \bibnamefont{and}
  \bibinfo{author}{\bibfnamefont{K.}~\bibnamefont{Ivanova}},
  \bibinfo{journal}{Phys. Chem. Earth B} \textbf{\bibinfo{volume}{24}},
  \bibinfo{pages}{781} (\bibinfo{year}{1999}).

\bibitem[{\citenamefont{Ivanova and Ausloos}(1999)}]{Ivanova14170}
\bibinfo{author}{\bibfnamefont{K.}~\bibnamefont{Ivanova}} \bibnamefont{and}
  \bibinfo{author}{\bibfnamefont{M.}~\bibnamefont{Ausloos}},
  \bibinfo{journal}{Physica A} \textbf{\bibinfo{volume}{274}},
  \bibinfo{pages}{349} (\bibinfo{year}{1999}).

\bibitem[{\citenamefont{Ivanova et~al.}(2000)\citenamefont{Ivanova, Ausloos,
  Clothiaux, and Ackermann}}]{Ivanova14160}
\bibinfo{author}{\bibfnamefont{K.}~\bibnamefont{Ivanova}},
  \bibinfo{author}{\bibfnamefont{M.}~\bibnamefont{Ausloos}},
  \bibinfo{author}{\bibfnamefont{E.}~\bibnamefont{Clothiaux}},
  \bibnamefont{and}
  \bibinfo{author}{\bibfnamefont{T.}~\bibnamefont{Ackermann}},
  \bibinfo{journal}{Europhys. Lett.} \textbf{\bibinfo{volume}{52}},
  \bibinfo{pages}{40} (\bibinfo{year}{2000}).

\bibitem[{\citenamefont{Ausloos and Ivanova}(2001)}]{Ausloos14150}
\bibinfo{author}{\bibfnamefont{M.}~\bibnamefont{Ausloos}} \bibnamefont{and}
  \bibinfo{author}{\bibfnamefont{K.}~\bibnamefont{Ivanova}},
  \bibinfo{journal}{Phys. Rev. E} \textbf{\bibinfo{volume}{6304}},
  \bibinfo{pages}{047201} (\bibinfo{year}{2001}).

\bibitem[{\citenamefont{Ivanova
  et~al.}(2002{\natexlab{a}})\citenamefont{Ivanova, Clothiaux, Shirer,
  Ackermann, Liljegren, and Ausloos}}]{Ivanova14140}
\bibinfo{author}{\bibfnamefont{K.}~\bibnamefont{Ivanova}},
  \bibinfo{author}{\bibfnamefont{E.}~\bibnamefont{Clothiaux}},
  \bibinfo{author}{\bibfnamefont{H.}~\bibnamefont{Shirer}},
  \bibinfo{author}{\bibfnamefont{T.}~\bibnamefont{Ackermann}},
  \bibinfo{author}{\bibfnamefont{J.}~\bibnamefont{Liljegren}},
  \bibnamefont{and} \bibinfo{author}{\bibfnamefont{M.}~\bibnamefont{Ausloos}},
  \bibinfo{journal}{J. Appl. Meteorology} \textbf{\bibinfo{volume}{41}},
  \bibinfo{pages}{56} (\bibinfo{year}{2002}{\natexlab{a}}).

\bibitem[{\citenamefont{Kitova et~al.}(2002)\citenamefont{Kitova, Ivanova,
  Ausloos, Ackermann, Clothiaux, and Mikhalev}}]{Kitova14130}
\bibinfo{author}{\bibfnamefont{N.}~\bibnamefont{Kitova}},
  \bibinfo{author}{\bibfnamefont{K.}~\bibnamefont{Ivanova}},
  \bibinfo{author}{\bibfnamefont{M.}~\bibnamefont{Ausloos}},
  \bibinfo{author}{\bibfnamefont{T.}~\bibnamefont{Ackermann}},
  \bibinfo{author}{\bibfnamefont{E.}~\bibnamefont{Clothiaux}},
  \bibnamefont{and} \bibinfo{author}{\bibfnamefont{M.}~\bibnamefont{Mikhalev}},
  \bibinfo{journal}{Int. J. Mod. Phys. C} \textbf{\bibinfo{volume}{13}},
  \bibinfo{pages}{217} (\bibinfo{year}{2002}).

\bibitem[{\citenamefont{Ivanova
  et~al.}(2002{\natexlab{b}})\citenamefont{Ivanova, Shirer, Clothiaux, Kitova,
  Mikhalev, Ackermann, and Ausloos}}]{Ivanova14120}
\bibinfo{author}{\bibfnamefont{K.}~\bibnamefont{Ivanova}},
  \bibinfo{author}{\bibfnamefont{H.}~\bibnamefont{Shirer}},
  \bibinfo{author}{\bibfnamefont{E.}~\bibnamefont{Clothiaux}},
  \bibinfo{author}{\bibfnamefont{N.}~\bibnamefont{Kitova}},
  \bibinfo{author}{\bibfnamefont{M.}~\bibnamefont{Mikhalev}},
  \bibinfo{author}{\bibfnamefont{T.}~\bibnamefont{Ackermann}},
  \bibnamefont{and} \bibinfo{author}{\bibfnamefont{M.}~\bibnamefont{Ausloos}},
  \bibinfo{journal}{Physica A} \textbf{\bibinfo{volume}{308}},
  \bibinfo{pages}{518} (\bibinfo{year}{2002}{\natexlab{b}}).

\bibitem[{\citenamefont{Ivanova and Ausloos}(2002)}]{Ivanova14110}
\bibinfo{author}{\bibfnamefont{K.}~\bibnamefont{Ivanova}} \bibnamefont{and}
  \bibinfo{author}{\bibfnamefont{M.}~\bibnamefont{Ausloos}},
  \bibinfo{journal}{J. Geophys. Res. - Atm.} \textbf{\bibinfo{volume}{107}},
  \bibinfo{pages}{4708} (\bibinfo{year}{2002}).

\bibitem[{\citenamefont{Ivanova et~al.}(2003)\citenamefont{Ivanova, Ackermann,
  Clothiaux, Ivanov, Stanley, and Ausloos}}]{Ivanova14100}
\bibinfo{author}{\bibfnamefont{K.}~\bibnamefont{Ivanova}},
  \bibinfo{author}{\bibfnamefont{T.}~\bibnamefont{Ackermann}},
  \bibinfo{author}{\bibfnamefont{E.}~\bibnamefont{Clothiaux}},
  \bibinfo{author}{\bibfnamefont{P.}~\bibnamefont{Ivanov}},
  \bibinfo{author}{\bibfnamefont{H.}~\bibnamefont{Stanley}}, \bibnamefont{and}
  \bibinfo{author}{\bibfnamefont{M.}~\bibnamefont{Ausloos}},
  \bibinfo{journal}{J. Geophys. Res. - Atm.} \textbf{\bibinfo{volume}{108}},
  \bibinfo{pages}{4268} (\bibinfo{year}{2003}).

\bibitem[{\citenamefont{Ausloos}(2004)}]{Ausloos14090}
\bibinfo{author}{\bibfnamefont{M.}~\bibnamefont{Ausloos}},
  \bibinfo{journal}{Physica A} \textbf{\bibinfo{volume}{336}},
  \bibinfo{pages}{93} (\bibinfo{year}{2004}).

\bibitem[{\citenamefont{Kurnaz}(2004{\natexlab{a}})}]{Kurnaz2004a}
\bibinfo{author}{\bibfnamefont{M.~L.} \bibnamefont{Kurnaz}},
  \bibinfo{journal}{J. Stat. Mech.: Theory and Experiment} p.
  \bibinfo{pages}{P07009} (\bibinfo{year}{2004}{\natexlab{a}}).

\bibitem[{\citenamefont{Kurnaz}(2004{\natexlab{b}})}]{Kurnaz2004b}
\bibinfo{author}{\bibfnamefont{M.~L.} \bibnamefont{Kurnaz}},
  \bibinfo{journal}{Fractals} p. \bibinfo{pages}{365}
  (\bibinfo{year}{2004}{\natexlab{b}}).

\bibitem[{\citenamefont{Mandelbrot and
  Wallis}(1969{\natexlab{a}})}]{Mandelbrot12180}
\bibinfo{author}{\bibfnamefont{B.~B.} \bibnamefont{Mandelbrot}}
  \bibnamefont{and} \bibinfo{author}{\bibfnamefont{J.~R.}
  \bibnamefont{Wallis}}, \bibinfo{journal}{Water Resources Research}
  \textbf{\bibinfo{volume}{5}}, \bibinfo{pages}{321}
  (\bibinfo{year}{1969}{\natexlab{a}}).

\bibitem[{\citenamefont{Mandelbrot and
  Wallis}(1969{\natexlab{b}})}]{Mandelbrot12190}
\bibinfo{author}{\bibfnamefont{B.~B.} \bibnamefont{Mandelbrot}}
  \bibnamefont{and} \bibinfo{author}{\bibfnamefont{J.~R.}
  \bibnamefont{Wallis}}, \bibinfo{journal}{Water Resources Research}
  \textbf{\bibinfo{volume}{5}}, \bibinfo{pages}{967}
  (\bibinfo{year}{1969}{\natexlab{b}}).

\bibitem[{\citenamefont{Arneodo et~al.}(1995)\citenamefont{Arneodo, Bacry,
  Graves, and Muzy}}]{Arneodo5250}
\bibinfo{author}{\bibfnamefont{A.}~\bibnamefont{Arneodo}},
  \bibinfo{author}{\bibfnamefont{E.}~\bibnamefont{Bacry}},
  \bibinfo{author}{\bibfnamefont{P.~V.} \bibnamefont{Graves}},
  \bibnamefont{and} \bibinfo{author}{\bibfnamefont{J.~F.} \bibnamefont{Muzy}},
  \bibinfo{journal}{Phys. Rev. Lett.} \textbf{\bibinfo{volume}{74}},
  \bibinfo{pages}{3293} (\bibinfo{year}{1995}).

\bibitem[{\citenamefont{Peng et~al.}(1994)\citenamefont{Peng, Buldyrev, Havlin,
  Simmons, Stanley, and Goldberger}}]{Peng13540}
\bibinfo{author}{\bibfnamefont{C.~K.} \bibnamefont{Peng}},
  \bibinfo{author}{\bibfnamefont{S.~V.} \bibnamefont{Buldyrev}},
  \bibinfo{author}{\bibfnamefont{S.}~\bibnamefont{Havlin}},
  \bibinfo{author}{\bibfnamefont{M.}~\bibnamefont{Simmons}},
  \bibinfo{author}{\bibfnamefont{H.~E.} \bibnamefont{Stanley}},
  \bibnamefont{and} \bibinfo{author}{\bibfnamefont{A.~L.}
  \bibnamefont{Goldberger}}, \bibinfo{journal}{Phys. Rev. E}
  \textbf{\bibinfo{volume}{49}}, \bibinfo{pages}{1685} (\bibinfo{year}{1994}).

\bibitem[{\citenamefont{Fraedrich and Blender}(2004)}]{Fraedrich12720}
\bibinfo{author}{\bibfnamefont{K.}~\bibnamefont{Fraedrich}} \bibnamefont{and}
  \bibinfo{author}{\bibfnamefont{R.}~\bibnamefont{Blender}},
  \bibinfo{journal}{Phys. Rev. Lett.} \textbf{\bibinfo{volume}{92}},
  \bibinfo{pages}{039802} (\bibinfo{year}{2004}).

\bibitem[{\citenamefont{Blender and Fraedrich}(2003)}]{Blender13260}
\bibinfo{author}{\bibfnamefont{R.}~\bibnamefont{Blender}} \bibnamefont{and}
  \bibinfo{author}{\bibfnamefont{K.}~\bibnamefont{Fraedrich}},
  \bibinfo{journal}{Geophys. Res. Lett.} \textbf{\bibinfo{volume}{30}},
  \bibinfo{pages}{1769} (\bibinfo{year}{2003}).

\bibitem[{\citenamefont{Cook}(1995)}]{Cook13050}
\bibinfo{author}{\bibfnamefont{R.}~\bibnamefont{Cook}},
  \bibinfo{journal}{Climate Dynamics} \textbf{\bibinfo{volume}{11}},
  \bibinfo{pages}{211} (\bibinfo{year}{1995}).

\bibitem[{\citenamefont{Bryson and Murray}(1977)}]{Bryson13060}
\bibinfo{author}{\bibfnamefont{E.~R.} \bibnamefont{Bryson}} \bibnamefont{and}
  \bibinfo{author}{\bibfnamefont{T.~J.} \bibnamefont{Murray}},
  \emph{\bibinfo{title}{Climates of Hunger}} (\bibinfo{publisher}{University of
  Wisconsin Press, Madison}, \bibinfo{year}{1977}).

\bibitem[{\citenamefont{Tan et~al.}(2003)\citenamefont{Tan, Liu, Hou, Qin,
  Zhang, and Li}}]{Tan}
\bibinfo{author}{\bibfnamefont{M.}~\bibnamefont{Tan}},
  \bibinfo{author}{\bibfnamefont{T.~S.} \bibnamefont{Liu}},
  \bibinfo{author}{\bibfnamefont{J.}~\bibnamefont{Hou}},
  \bibinfo{author}{\bibfnamefont{X.}~\bibnamefont{Qin}},
  \bibinfo{author}{\bibfnamefont{H.}~\bibnamefont{Zhang}}, \bibnamefont{and}
  \bibinfo{author}{\bibfnamefont{T.}~\bibnamefont{Li}},
  \bibinfo{journal}{Geophysical Research Letters}
  \textbf{\bibinfo{volume}{30}}, \bibinfo{pages}{1617} (\bibinfo{year}{2003}).

\bibitem[{\citenamefont{Chase-Dunn et~al.}(2006)\citenamefont{Chase-Dunn,
  Niemeyer, Alvarez, Inoue, Lawrence, and Carlson}}]{Chase-Dunn}
\bibinfo{author}{\bibfnamefont{C.}~\bibnamefont{Chase-Dunn}},
  \bibinfo{author}{\bibfnamefont{R.}~\bibnamefont{Niemeyer}},
  \bibinfo{author}{\bibfnamefont{A.}~\bibnamefont{Alvarez}},
  \bibinfo{author}{\bibfnamefont{H.}~\bibnamefont{Inoue}},
  \bibinfo{author}{\bibfnamefont{K.}~\bibnamefont{Lawrence}}, \bibnamefont{and}
  \bibinfo{author}{\bibfnamefont{A.}~\bibnamefont{Carlson}}
  (\bibinfo{year}{2006}).

\bibitem[{\citenamefont{Koeppen}(1900)}]{Koeppen13800}
\bibinfo{author}{\bibfnamefont{W.}~\bibnamefont{Koeppen}},
  \bibinfo{journal}{Geographischen Zeitschrift} \textbf{\bibinfo{volume}{9}},
  \bibinfo{pages}{593} (\bibinfo{year}{1900}).

\bibitem[{\citenamefont{Vapnik}(1995)}]{Vapnik13740}
\bibinfo{author}{\bibfnamefont{V.}~\bibnamefont{Vapnik}},
  \emph{\bibinfo{title}{The Nature of Statistical Learning Theory}}
  (\bibinfo{publisher}{Springer Verlag}, \bibinfo{year}{1995}).

\bibitem[{\citenamefont{Joachims}(1999)}]{Joachims13750}
\bibinfo{author}{\bibfnamefont{T.}~\bibnamefont{Joachims}},
  \emph{\bibinfo{title}{Advances in Kernel Methods - Support Vector Learning}}
  (\bibinfo{publisher}{M.I.T. Press}, \bibinfo{year}{1999}).

\bibitem[{\citenamefont{FluegemanJr and Snow}(1989)}]{Fluegeman9510}
\bibinfo{author}{\bibfnamefont{R.~H.} \bibnamefont{FluegemanJr}}
  \bibnamefont{and} \bibinfo{author}{\bibfnamefont{R.~S.} \bibnamefont{Snow}},
  \bibinfo{journal}{Pageoph.} \textbf{\bibinfo{volume}{131}},
  \bibinfo{pages}{307} (\bibinfo{year}{1989}).

\bibitem[{\citenamefont{Wang et~al.}(1992)\citenamefont{Wang, Evans, Xu,
  Rutter, Ding, and Liu}}]{Wang7920}
\bibinfo{author}{\bibfnamefont{Y.}~\bibnamefont{Wang}},
  \bibinfo{author}{\bibfnamefont{M.~E.} \bibnamefont{Evans}},
  \bibinfo{author}{\bibfnamefont{T.~C.} \bibnamefont{Xu}},
  \bibinfo{author}{\bibfnamefont{N.}~\bibnamefont{Rutter}},
  \bibinfo{author}{\bibfnamefont{Z.}~\bibnamefont{Ding}}, \bibnamefont{and}
  \bibinfo{author}{\bibfnamefont{X.~M.} \bibnamefont{Liu}},
  \bibinfo{journal}{Can. J. Earth Sci.} \textbf{\bibinfo{volume}{29}},
  \bibinfo{pages}{296} (\bibinfo{year}{1992}).

\end{thebibliography}
\end{document}